\documentclass[
twocolumn,
]{ceurart}

\usepackage{amsmath}
\usepackage{rotating}
\newcommand{\etal}{\textit{et al.}}
\usepackage{color}
\usepackage{colortbl}
\usepackage{listings}
\begin{document}

\graphicspath{{images/}}

\copyrightyear{2023}
\copyrightclause{Copyright for this paper by its authors.
  Use permitted under Creative Commons License Attribution 4.0
  International (CC BY 4.0).}

\conference{Joint Proceedings of the ACM IUI Workshops 2023, March 2023, Sydney,
Australia,}

\title{Distinguishing Engagement Facets: An Essential Component for AI-based Interactive Healthcare}


\author[1,2]{Hanan Salam}[%
orcid=0000-0001-6971-5264,
email=hanan.salam@nyu.edu,
url=https://wp.nyu.edu/smartlab/,
]
\cormark[1]
\fnmark[1]
\address[1]{New York University Abu Dhabi, PO Box 129188, Saadiyat Island, Abu Dhabi, United Arab Emirates}
\address[2]{Social Machines \& Robotics (SMART) Lab, Center of AI \& Robotics (CAIR)}



\cortext[1]{Corresponding author.}

\begin{abstract}
  Engagement in Human-Machine Interaction is the process by which entities participating in the interaction establish, maintain, and end their perceived connection. It is essential to monitor the engagement state of patients in various AI-based interactive healthcare paradigms. This includes medical conditions that alter social behavior such as Autism Spectrum Disorder (ASD) or Attention-Deficit/Hyperactivity Disorder (ADHD). Engagement is a multi-faceted construct which is composed of behavioral, emotional, and mental components. Previous research has neglected this multi-faceted nature of engagement and focused on the detection of engagement level or binary engagement label. In this paper, a system is presented to distinguish these facets using contextual and relational features. This  can facilitate further fine-grained analysis. Several machine learning classifiers including traditional and deep learning models  are compared for this task. An F-Score of $0.74$  was obtained on a balanced dataset of 22242 instances with neural network-based classification. The proposed framework shall serve as a baseline for further research on engagement facets recognition, and its integration is socially assistive robotic applications.
\end{abstract}


\maketitle

\section{Introduction}
\label{sec:intro}

During the last decade, researchers have demonstrated interest in enhancing the capabilities of robots to assist humans in their daily life. 
This requires incorporation of social intelligence within the robots which involves understanding different states of engagement. 

Research in Human-Machine Interaction (HMI) has depicted that engagement is a multi-faceted construct and consists of different components \cite{salam2022automatic}.
It is very much important to be able to distinguish the facets before performing a deeper analysis.
Corrigan~\etal~\cite{Corrigan2016} demonstrated that engagement is mainly composed of \textit{cognitive} and \textit{affective} components which are manifested by attention and enjoyment.  
According to O'brien~\etal~\cite{o2008user}, engagement is characterized by features like \textit{challenge}, \textit{positive affect}, \textit{endurability}, \textit{aesthetic and sensory appeal}, \textit{attention}, \textit{feedback}, \textit{variety/novelty}, \textit{interactivity}, and \textit{perceived user control}. In the context of youth engagement in activities, Ramey~\etal~\cite{ramey2015measuring} proposed a model of psychological engagement having three components: \textit{cognitive} like thinking or concentrating, \textit{affective} like enjoyment, and \textit{relational} like through connectedness to something. Salam \etal~\cite{salam2015multi} showed that the mental and emotional states of the user related to engagement vary in function of the current interaction context.
These studies suggest that when attempting automatic inference of user's engagement state, it is important to consider this multi-faceted nature.

Application areas of Assistive Robotics include elderly care \cite{broadbent2011human}, helping people with medical conditions that alter social behavior such as children suffering from Autism Spectrum Disorder (ASD) \cite{feil2009development} or people suffering from Adult Deficient Hyperactivity Disorder (ADHD)~\cite{fridin2011educational}, coaching and tutoring \cite{greczek2014expanding,zhu2012face}. 
Fasola and Mataric~\cite{fasola2013socially} presented a Socially Assistive Robot (SAR) system designed to engage elderly users in physical exercise. Different variants of the robot's verbal instructions were used to minimize the robot's perceived verbal repetitiveness, and thus maintain the users' engagement. Previous engagement detection approaches revolve around a binary classification-based approach (engaged vs. not engaged) \cite{foster2013can,leite2015comparing} or a multi-class approach (engagement level) \cite{castellano2012detecting,peters2010investigating}. However, the multi faceted nature is seldom considered. 

In this paper, a framework that takes into account the multi-faceted nature of engagement is proposed.  
Engagement is modeled in terms of a spectrum of engagement states: mental, behavioural and emotional. 
This is the first engagement framework of its kind to propose such classification framework of the facets of engagement. Such analysis allows to inform the implementation of fine-grained strategies based on a deeper understanding of user's states. 
We present a preliminary evaluation of this approach on an off-line multi-party HRI corpus. The corpus was chosen due to the relevance of  its interaction scenario  (educational followed by competitive context) to the use of AI-based interactive healthcare systems. For instance, SAR for neuro-developmental disorders such as ADHD and ASD, which might benefit from a multi-faceted engagement model. For instance, an educational scenario can be adopted for the characterization of ADHD, since such context would solicit attention cues which are normally impaired in the case of ADHD individuals. 
We are aware that for the study to be complete, it should be validated in the context of an SAR scenario. However, the lack of such dataset has led us to choose a proxy dataset to perform the initial validation of the framework.


\section{Related Work}
\label{sec:relatedWork}

Engagement in Human-Robot Interaction is defined as the process by which two (or more) participants establish, maintain, and end their perceived connection \cite{sidner2004look,sidner2005explorations}. Andrist \etal~\cite{andrist2017went} analyzed an HRI dataset in terms of interaction type, quality, problem types, and the system's  failure points causing problems. Failure in the engagement component was found to be among the major identified problems cause during the interaction. This confirms that a highly performing engagement model is essential for the success of any HRI scenario  \cite{andrist2017went}.

Bohus \& Horvitz~\cite{bohus2009models} pioneered research on engagement in multi-party interaction. They explored disparate engagement strategies to allow robots to engage simultaneously with multiple users. 
There are multi-farious studies based on multi-party interactions.
Oertel~\etal~\cite{oertel2013gaze} studied in both individual and group level  about the relationship amidst the participants' gaze and speech behavior. 
Leite~\etal~\cite{leite2015comparing} experimented with the generalization capacity of an engagement model. It was trained and tested on single-party and multi-party scenarios respectively. The opposite scenario was also considered. 
Salam \etal~\cite{salam2015engagement} conducted a study on engagement recognition in a triadic HRI scenario and showed that it is possible to infer a participant's engagement state based on  the other participants' cues.

Most of engagement inference approaches revolved around identification of a person's intention to engage. There has also been studies to detect whether the person is engaged/disengaged. Benkaouar \etal~\cite{benkaouar2012multi} presented a system to detect disparate engagement phases. This includes intention to engage, engaged and disengaged. Foster \etal~\cite{foster2013can} attempted to detect whether a person intends to engage which is a bi-class problem. Leite \etal~\cite{leite2015comparing} attempted to identify disengagement in both group and individual interactions. Ben \etal~\cite{ben2019fly} also presented a system dedicated to the similar cause.
 
There are different works which focused on detecting different levels of engagement of a user. Michalowski \etal~\cite{michalowski2006spatial} distinguished different levels of engagement in the thick of present, interacting, engaged, and just attending.  A system to distinguish two classes of engagement namely medium-high to high and medium-high to low engagement was presented by~\cite{castellano2012detecting}. MBednarik \etal~\cite{Bednarik:2012} distinguished disparate states of conversational engagement states. It includes no interest, following, responding, conversing, influencing  and managing. They also modelled a bi-class problem having low/high conversational level. Oetrel~\etal~\cite{oertel2013gaze} distinguished 4 classes for group involvement namely high, low, leader, steering the conversation, and group is forming itself. Two models were developed in \cite{dewan2018deep} focusing on not-engaged/engaged and not-engaged/normally-engaged/very-engaged state distinction. Frank \etal~\cite{frank2016engagement} differentiated 6 different states of engagement in the thick of disengagement, involved engagement, relaxed engagement, intention to act, action, and involved action.

Recently, \cite{devillers2018multifaceted} stated that engagement in HRI should be multi-faceted. Formulating binary/multi-class problems (engaged vs. not engaged) or a multi-class problem (engagement level) over this ignore the multi-faceted nature of engagement. Taking this multi-faceted nature  into consideration is very important for the design of intelligent social agents. For instance, this can influence the implemented engagement strategies within the agent's architecture. 
Some studies attempted to implement different strategies related to task and social engagement. For instance, \cite{el2019learning} implemented a task engagement strategy which focuses on the task at hand and having users meta-cognitively reflect on the robot's performance and a social engagement strategy which focuses on their enjoyment and having them meta-cognitively reflect on their emotions with respect to the activity and the group interactions.


Different features have been used to distinguish engagement states. Some of such features include contextual ~\cite{kapoor2004probabilistic,castellano2012detecting,salam2015engagement} attentional~\cite{Yun2012,Papadopoulos2013}, affective~\cite{castellano2012detecting,foster2013can,dewan2018deep,masui2020measurement} to name a few. Salam \etal~\cite{salam2016fully} used person to detect both individual and group engagement. \cite{inoue2018engagement,inoue2019latent} combined different aspects like backchannels, eye gaze, head nodding-based features to detect engagement level. Ben \etal~\cite{ben2019fly} combined several attributes like speech and facial expressions, gaze and head motion, distance to robot to identify disengagement. Masui \etal~\cite{masui2020measurement} worked with facial Action Units and physiological responses.

Recent approaches explored deep learning architectures for the detection of engagement. Dewan \etal~\cite{dewan2018deep} used person-independent edge features and Kernel Principal Component Analysis (KPCA) within a deep learning framework to detect online learners' engagement using facial expressions. \cite{del2020you} used CNN and LSTM networks to predict engagement level. \cite{chithrra2022personalized} proposed adaptive deep architectures for different user groups for predicting engagement in robot-mediated collaborative learning.


Contextual information is being used in social signal processing for quite some time.
Kapoor~\etal~\cite{kapoor2005multimodal} combined context features in the form of game state with facial and posture features in an online educative scenario. Martinez and Yannakakis~\cite{martinez2011mining} used sequence mining for the prediction of computer game player affective states.
Castellano~\etal~\cite{castellano2012detecting} explored task and social-based contextual features.   
In another instance, the authors~\cite{castellano2016detecting} used same contextual features for distinguishing interaction quality.

Relational feature have proven to be useful in multifarious instances. Curhan \etal~\cite{curhan2007thin} used dyad-based cues for predicting negotiation outcomes. 
Jayagopi \etal~\cite{jayagopi2012linking} adhered to group-based cues to understand typical behavior in small groups. Nguyen \etal~\cite{nguyen2013hire} extracted relational audio-visual cues to detect the suitability of an applicant in a job interview. The features included audio and visual back-channeling, nodding while speaking, mutual short utterances and nods. It also includes~\cite{biel2013youtube} that used ``looking-while-speaking'' feature to understand personality impressions from conversational logs extracted from YouTube.

So far, context has been insufficiently investigated in the avenue of affective and cognitive states. Devillers \etal~\cite{devillers2017toward} highlights the importance of context in the assessment of engagement. They identified paralinguistic, linguistic, non-verbal, interactional, and specific emotional and mental state-based features as very important for engagement prediction. In this work, we investigate relational and contextual features for the recognition of a spectrum of engagement states. The features have been used in isolation as well as in combination to assess their engagement state distinction capability. These features have not been combined previously for detecting engagement facets. Compared to previous works, the proposed features model interaction context, the robot's behavior and the behavioral relation between the participant in question and the other entities of the interaction.


\section{Need for Engagement Recognition in Interactive Healthcare}

Technological advancements have propagated to every field. There has always been efforts to automate tasks. Healthcare is one of the primal needs for society and it also has been touched by technology \cite{thelisson2018general}. Several interactive systems have come up to aid in automated healthcare and the well-being of people with medical conditions. In~\cite{fridin2011educational}, an SAR is proposed, whose aim is to help children with ADHD to improve their educational outcome through social interaction with a robot. Another educational SAR was presented by~\cite{greczek2014socially}. This was targeted towards providing assistance in personalizing education in classrooms. Children with Autism Spectrum Disorder are a target population for such personalized teaching systems \cite{kasari2018smarter}. However, most of existing systems do not include a user's engagement analysis module. Such Socially Assistive systems can largely benefit from a fine-grained analysis of engagement. This will make the systems more human-like.
There has been interest in automated screening and consultation to detect problems of the body and mind at an early stage. This can also help to reduce the initial load on doctors. It is very important for the patients to feel that they are interacting with their peers rather than a machine. The systems need to process both audio and visual cues in order to properly understand patients. While the patients are interacting with the automated systems, several states of engagement needs to be monitored simultaneously. This includes level of concentration, different reactions, spontaneity, to name a few. Such states of engagement portray useful information about a patient's health. These engagement states can be categorized into a broader spectrum of behavioral, mental and emotional states. Distinguishing the engagement facet is important at the outset for a deeper analysis. This can pave the way for systems which would better understand the condition of patients by reading their body language and wont merely match spoken symptoms. This will especially be useful in treating and understanding mental conditions where the body language is a vital aspect. In the case of psychological problems, patients are often engaged into conversations regarding disparate aspects by doctors wherein the patient's body language serves a vital pointer towards the mental condition.

\section{Proposed Framework}

\begin{figure*}[t!]
  \begin{centering}
    \includegraphics[scale=0.35]{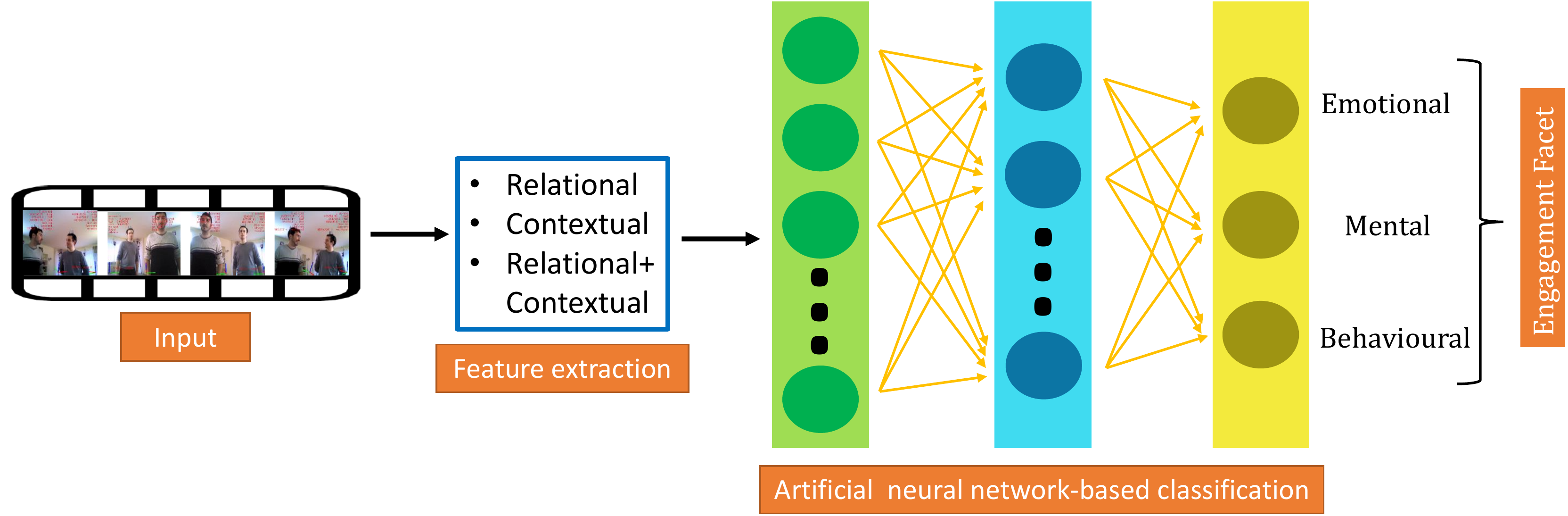}\\
  \end{centering}
  \caption{Illustration of the proposed framework with an Artificial Neural Network classifier.}
  \label{fig:bd}
\end{figure*}

The proposed framework is composed of 3 steps. First, a multi-party HRI corpus is annotated in terms of engagement facets. Then, different contextual and relational features are extracted. Finally, different standard classifiers are used to classify the different engagement facets. Fig. \ref{fig:bd} presents an illustration of the proposed framework.

\subsection{Data Corpus}
\label{sec:setup}

\begin{figure}[t!]
  \begin{centering}
    \includegraphics[scale=0.4]{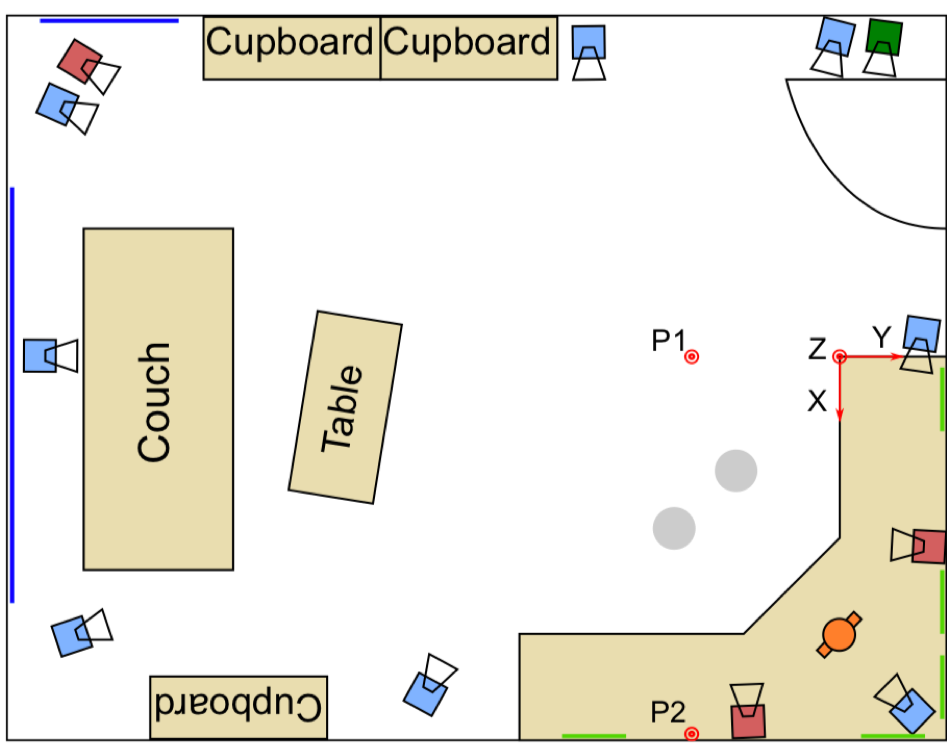}\\
  \end{centering}
 \caption{Organisation of the recording room. NAO (orange), participants typical positions (gray circles), cameras (HD: red, VICON: blue),  wizard feedback (green), paintings (green lines), windows (blue lines),  VICON coordinate system (red),  head pose calibration position(P1 and P2). }
  \label{fig:snapshotvernissage}
\end{figure}

In this section, the data corpus along with the disparate engagement annotations is discussed. 

\subsection{Interaction Scenario \& Modalities}

We use $4$ interactions of $8$ participants from the conversational  HRI data corpus `Vernissage' \cite{jayagopi2012vernissage}. It is a multi-party interaction amidst the humanoid robot NAO\footnote{https://www.softbankrobotics.com/emea/en/nao} and $2$ participants. The interaction has different contexts which can mainly be differentiated into $2$ parts. The $1^{st}$ is where the robot describes several paintings hanged on a wall (informative/educational context). In the $2^{nd}$ the robot performs a quiz with the volunteers related to art and culture (competitive context). This was done in order to encompass different variations for the engagement states.

This corpus was chosen since its interaction scenario is relevant to the use of SAR for neuro-developmental disorders such as ADHD and ASD, that might benefit from a multi-faceted engagement model. For instance, an educational scenario like the once in the first part of the Vernissage corpus scenario can be adopted for the characterization of ADHD, since an educational/informative scenario would solicit attention cues which are normally impaired in the case of ADHD individuals. 
We are aware that for the study to be complete, it should be validated in the context of an SAR scenario. However, the lack of such dataset has led us to choose a proxy dataset to perform our initial validation of the framework.

The average length per interaction is nearly $11$ minutes. 
NAO's internal camera was used to record the clips. This provided the front view. $3$ other cameras were also used to get the left, right and rear feeds.  Fig.~\ref{fig:snapshotvernissage} shows the organization of the recording room. 

The corpus has annotations for non-verbal behaviors of the participants. It also contains robot's speech and action in the log file of the robot.
\subsection{Engagement Annotations}
\label{sec:engAnn}

Engagement labels were assigned to $3$ categories namely mental, behavioral and emotional. These were annotated when the participants manifested one the following states: thinking, listening, positive/negative reaction, responding, waiting for feedback, concentrating, and listening to the other participant. The annotations were performed by $2$ people with the aid of Elan\footnote{https://tla.mpi.nl/tools/tla-tools/elan/} annotation tool~\cite{wittenburg2006elan}. They watched every video $2$ times (once with the perspective of $1$ participant). Discrete segments were annotated and it was stopped as soon as a change was observed.  The Mean inter-rater Cronbach' Alpha coefficient was $0.93$. This points to the reliability of the annotations. The details of each category is as follows.

\paragraph{\textbf{Mental states} --} A segment was assigned mental state label when the participant manifested one of the following mental states:

\begin{itemize}
\item \textit{Listening (EL)}: The participant is listening to NAO;

\item	\textit{WaitingFeedback (EWF)}: The participant is waiting for NAO's feedback after he/she had answered a question; 

\item	\textit{Thinking (ETh)}: The participant is thinking about the response to a question asked by NAO; 

\item	\textit{Concentrating (EC)}: The participant is concentrating with NAO; 

\item	\textit{ListeningPerson2 (ELP2)}: The participant is listening to the other who is answering NAO.

\end{itemize}

\paragraph{\textbf{Behavioral states} --} A segment was assigned a behavioral state label when the participant manifested the following behavioral state:
\begin{itemize}
\item	\textit{Responding (ER)}: The participant is responding to NAO;
\end{itemize}

\paragraph{\textbf{Emotional states} --} A segment was assigned an emotional state label when the participant manifested one of the following emotional states:
\begin{itemize}
\item	\textit{PositiveReaction (EPR)}: The participant shows a positive reaction to  NAO. 

\item	\textit{NegativeReaction (ENR)}: The participant shows a negative reaction to NAO. 

\end{itemize}
The details regarding the number of annotated instances for each class is presented in Table ~\ref{tab:NumInstances}.

\begin{table}
\caption{Details on the number of annotated instances in each class}
\label{tab:NumInstances}
\centering
\begin{tabular}{|l||c|}
\hline
\textbf{State}&\textbf{Number of instances}\\
\hline \hline

Behavioral        &10331\\

Emotional       &7414\\

Mental       &80902\\

    \hline
\hline
\textbf{Total}	& 98647\\
\hline
\end{tabular}
\end{table}

%

\subsection{Extracted Features}
\label{sec:featExt}

In this study, we used the annotated cues from Vernissage corpus. Moreover, we extracted additional metrics which were computed from the existing ones. They were categorized into two categories: 1) \textit{contextual} and 2) \textit{relational}. 

Contextual features deal with either the different entities of an interaction like the robot utterance, addressee and topic of speech or behavioral aspects of the participant that concern the interaction context like visual focus of attention and addressee. 

\begin{figure}[htbp]
	\centering
		\includegraphics[scale=0.3]{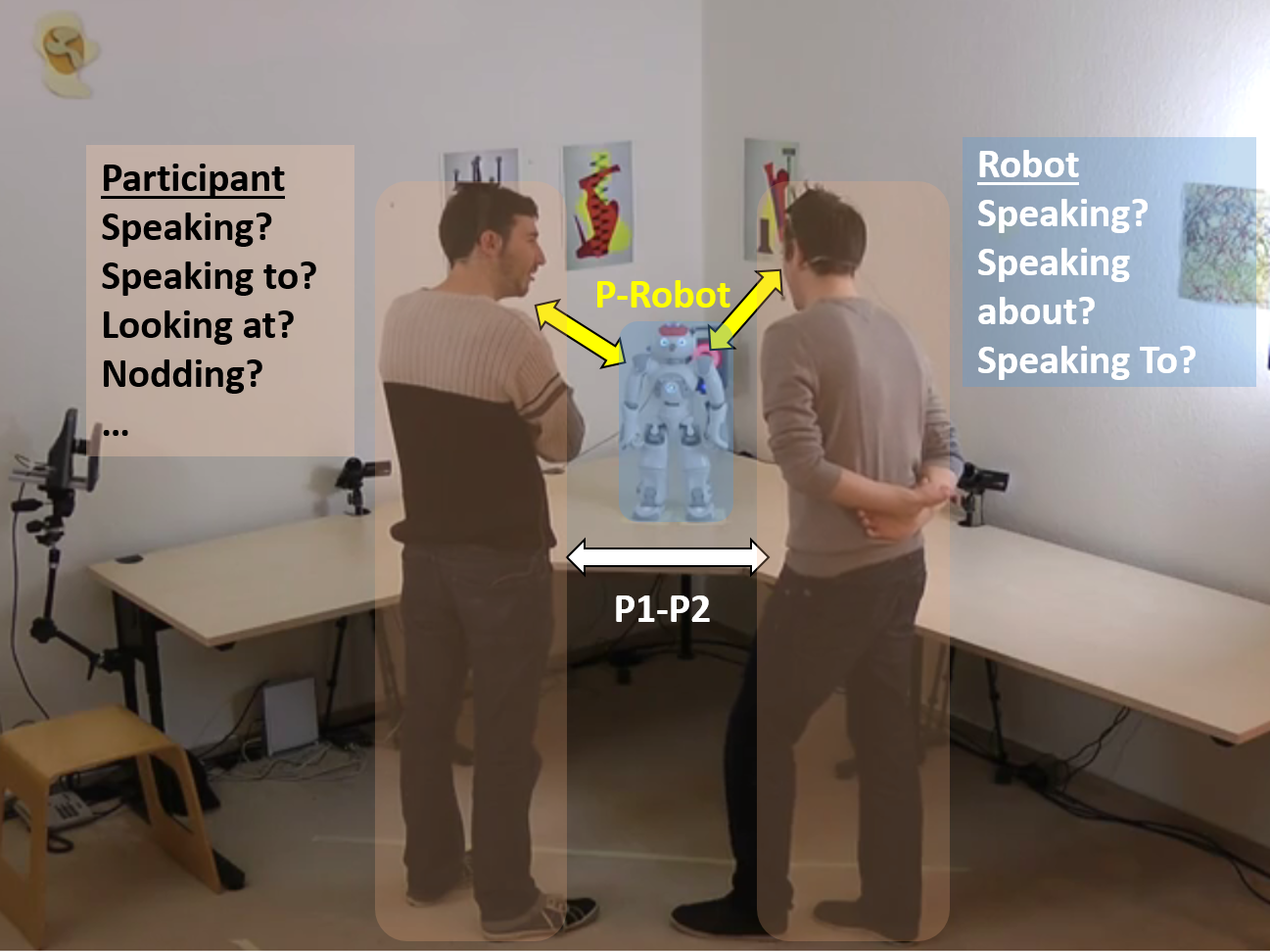}
		\caption{Features illustration: contextual features (Robot, Participant); relational features (Participant-Robot, Participant1-Participant2)}
		\label{fig:featuresIllustration}
\end{figure}

Relational features encode the behavioral relation between the participants and the robot. Fig.~\ref{fig:featuresIllustration} illustrates the features groups used in our study.

\subsubsection{Contextual Features}
Interaction amidst entities involves both entities and connection. While inferring the engagement state of an interacting person, we consider behavior of the person as well as our behavior. 
Thus, an automated engagement identification system should also consider the same. 

Consequently, we employ different contextual features that describes the participant's behavior with respect to the other entities.
Moreover, for a dialogue of the robot, we extract the robot's utterance, addressee and topic of speech.

\paragraph{Participant:}
\begin{enumerate}
\item \textit{Visual Focus Of Attention (VFOA)}:  Gaze in human-human social interactions  is considered as the primary cue of attention \cite{mason2005look,sidner2003engagementwhenlooking}.  
We use VFOA ground truth of every participant which were annotated with $9$ labels. 

\item \textit{VFOA Shifts}: 
Gaze shifts  indicate people's engagement/ disengagement
with specific environmental stimuli \cite{baron1997mindblindness}. We define VFOA shift as the moment when a participant shifts attention to a different subject. This feature is binary and is computed from the VFOA labels.

\item \textit{Addressee}: When addressing  somebody, we are engaged with him/her. Similarly, in the context of HRI, when a participant addresses someone other than the robot, he/she is disengaged from the robot. Adressee annotations  used from the corpus and are annotated into 6 Classes: \{NoLabel, Nao, Group, PRight, PLeft, Silence\}.
\end{enumerate}

\paragraph{Robot:}
Starting from the robot's conversation logs, the following were extracted. 
\begin{enumerate}
\item \textit{Utterances}: The labels \{Speech, Silence\} were assigned to frames depending on the robot's speech activity.
\item \textit{Addressee}: The addressee of the robot was detected using predefined words from its speech. The following labels were assigned \{Person1, Person2, GroupExplicit, GroupPerson1, GroupPerson2, Person1Group, Person2Group, Group, Silence\}. `GroupExplicit' label refers to such segments where the robot was explicitly addressing both participants. `GroupPersonX' $/ X\in (1,2)$ corresponds to segments where the robot addresses the group then `PersonX' while `PersonXGroup' represents the inverse. 

\item \textit{Topic of Speech}: This was identified using a keyword set. These were related to disparate paintings available in the scene \{manray, warhol, arp, paintings\}. Frames were  allotted labels based on them. 

\end{enumerate}

\subsubsection{Relational Features}

\begin{figure}[htbp]
	\centering
		\includegraphics[scale=0.28]{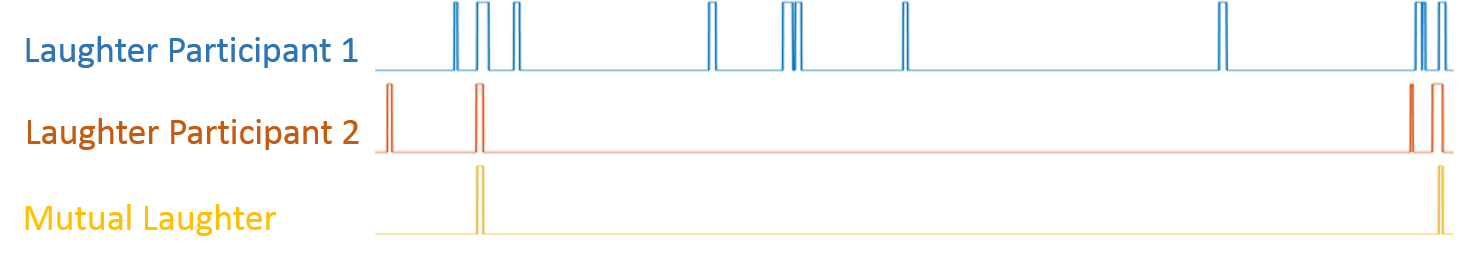}
		\caption{Example of relational cues extraction. This corresponds to participants' mutual laughter detection using logical AND from laughter time series.}
		\label{fig:mutualLaugh}
\end{figure}

We extract a set of Relational Features describing robot's and participants' behaviors synchrony and alignment. These include, among others, mutual gaze and laughter. 
A logical AND operation was used between participants' and robot's features time series for obtaining mutual events occurrence. Fig.~\ref{fig:mutualLaugh} shows an example of participants' mutual laughter extraction.

\paragraph{Participant-Robot Features:}
\begin{enumerate}
\item \textit{Gaze-Speech Alignment}:
We extracted events where a participant looks at objects corresponding to the robot's topic of speech. 
This indicates that the participant is listening to the robot and is interested in what it is saying.
\item \textit{P1 talks to P2/Robot Speaks}: This refers to events where the participants speak with each other during the robot's speech. This may signal a disengagement behavior.
\end{enumerate}

\paragraph{Person1-Person2 Features:}
\begin{enumerate}
\item\textit{Participants Mutual Looks}: This refers to events where the participants look at each other. Though this may signal disengagement but it may also signal engagement as it might be a reaction to the robot's speech. 

\item \textit{Participants Mutual Laughter}: This refers to events where the two participants laugh together. This represents reaction to the robot's speech. 

\item \textit{P1 Looks at P2/ P2 Talks to Robot}: This represents  events where the passive participant looks at  active participant while he/she is talking to the robot. Though this may appear to be disengagement, but analysis revealed the inverse. 
\end{enumerate}

The total number of features is $39$. There were 34 contextual features and 5 relational features. 


\subsection{Engagement Facets Classification}
\label{sec:classification}
As this is the first work that proposed the classification of engagement facets, namely, behavioural, emotional, and mental, it is important to establish a classification baseline.  

We compare different classifiers for the defined engagement facets classification. 
The classifiers include traditional machine learning classifiers such as Bayesian Network (Bayes Net), Naive Bayes, Linear Logistic Regression (LLR), Support Vector Machine (SVM), Radial Basis Function Network (RBF Net), and simple Artificial Neural Network (ANN). A deep learning classifier, namely Recurrent Neural Network (RNN) was also used for this classification task. This helps establish an initial understanding of whether traditional machine learning classifiers are sufficient for the task, or more sophisticated classification techniques such as deep learning methods are needed.

\section{Results and Discussion}

The proposed framework is evaluated using $5$-fold cross validation.
As the data was highly imbalanced, so a subset of the data was drawn having equal number of instances per class totalling to $22242$ instances.  
The combined features (contextual+relational) for this dataset were used to train the different classifiers presented in section \ref{sec:classification}. 

\subsection{Comparative performance analysis of standard classifiers}
Table \ref{tcomp} presents the results of training different classifiers on the combined features. 
From the table, we can state that the best performing classifier was ANN with an accuracy of $74.57 \%$, followed by the Linear Logistic Regression model ($70.35 \%$). The performance of SVM and Bayes Net were very close ($69.6 \%$), followed by Naive Bayes ($68.61 \%$). Surprisingly, the least accuracy of $57.68\%$ was obtained using the deep RNN. This might be due to the fact that the number of samples is not sufficient for training deep neural networks. Consequently, traditional Machine Learning approaches performed better. It might be worth it to investigate deep neural networks in future works using a higher number of instances.

\begin{table}[]
\centering
\caption{Comparative analysis of the performance of standard classifiers on the balanced dataset.}
\label{tcomp}
\begin{tabular}{|l|>{}c|}
\hline
\textbf{Classifier} &\textbf{Accuracy} \textbf{(\%)}\\
                             \hline
\textbf{RNN}  &57.68 \\ 
\textbf{RBF Net}  & 65.13\\ 
\textbf{Naive Bayes}  & 68.61 \\ 
\textbf{Bayes Net}  &69.61  \\ 
\textbf{SVM}  & 69.68\\ 
\textbf{LLR}  &70.35 \\ 
\textbf{ANN}  &\textbf{74.57}\\ \hline
\end{tabular}
\end{table}

\subsection{Performance analysis on engagement facets}
We analyse the performance of the best performing classifier (ANN) on the different engagement facets (behavioral, emotional, mental). 
The corresponding confusion matrix is presented in Table \ref{balancedperf}. The values for different performance metrics (true positive rate, false positive rate, precision, recall, and F-score) for each of the classes are also presented in Table \ref{balpermet}. 

It is noted that the best performance was obtained for the behavioral class with an F-score of $0.794$. This was followed by the mental class where an F-score of $0.730$ was obtained. The lowest performance was obtained for the emotional class with an F-score of $0.707$. This lower performance for the emotional class might be explained by the fact that the current features might not be highly correlated with the emotional states, and more correlated with the other engagement facets. It might be worth it to investigate other relevant features in the future. 

Looking at the confusion matrix, we can see that the highest confused pair was behavioral-emotional where $1318$ instances were predicted as behavioral when they actually belonged to the emotional class. This is followed by the mental-emotional pair with $1203$ instances mis-classified as mental when their actual label was emotional. Similarly, $1153$ mental instances were mis-classified as emotional. The high confusion between mental and emotional engagement states is expected as these states might exhibit similar non-verbal cues. The confusion between behavioral and emotional states is less evident. This confusion might be due to the used features, which are not sufficient to precisely predict the emotional states.

\begin{table}[h!t]
\centering
\caption{Confusion matrix for balanced dataset.}
\label{balancedperf}
\begin{tabular}{|>{}l||>{}c|>{}c|>{}c|}
\hline
           & \textbf{Behavioral} & \textbf{Emotional} & \textbf{Mental} \\ 
           \hline
              \hline
\textbf{Behavioral} & 6347       & 1318       & 916   \\ \hline
\textbf{Emotional}  & 384        & 4893      & 1153   \\ \hline
\textbf{Mental}     & 683        & 1203       & 5345  \\ \hline
\end{tabular}
\end{table}

\begin{table}[h!t]
\centering
\caption{Class-wise values for performance metrics on the balanced dataset. True Positive Rate (TPR), False Positive Rate (FPR).}
\label{balpermet}
\begin{tabular}{|>{}l||>{}c|>{}c|>{}c|}
\hline
    \textbf{Metrics}       & \textbf{Behavioral} & \textbf{Emotional} & \textbf{Mental} \\ 
           \hline
              \hline
\textbf{TPR} & 0.856       & 0.660       & 0.721   \\ \hline
\textbf{FPR}  & 0.151& 0.104     & 0.127   \\ \hline
\textbf{Precision}     & 0.740        & 0.761       & 0.739  \\ \hline
\textbf{Recall}  & 0.856        & 0.660      & 0.721   \\ \hline
\textbf{F-score}  & 0.794        & 0.707      & 0.730   \\ \hline
\end{tabular}
\end{table}


\section{Conclusions and future work}
\label{sec:conc}

In this paper, we proposed a system to detect different facets of engagement states: mental, emotional, and behavioral. This is the first engagement framework of its kind to propose such classification framework of engagement. This is essential for a deeper analysis of the user's engagement by machines. In the context of AI-based  healthcare systems such as socially assistive robots, such fine-grained analysis would improve performance, and facilitate adaptive interventions. For instance, recognizing whether the user's engagement is emotional, behavioral, or mental, might better inform AI-based healthcare systems, especially those that rely on interactive systems (e.g. ASR for ADHD or ASD).
The proposed framework was validated on an HRI corpus exhibiting educational and competitive contexts, which are relevant to AI-based interactive systems.  The preliminary results show that it is possible to classify engagement facets with a relatively acceptable accuracy. These results shall serve as a baseline for the development of more accurate systems. 
In future, we plan to validate the framework on a larger dataset that exhibits an SAR scenario.  We plan to work with individual features to improve the system's performance and perform a deep grained analysis of the different states. We will also explore deep learning-based approaches and unsupervised approaches towards detection of engagement state types and thereafter finer classification. Deep learning will be used not only for data classification but also for feature extraction.

\begin{acknowledgments}
This work is supported in part by the NYUAD Center for Artificial Intelligence and Robotics, funded by Tamkeen under the NYUAD Research Institute Award CG010.
\end{acknowledgments}

\bibliography{IJSR_SALAM}

\end{document}